\documentclass{article}

\usepackage{CJKutf8}
\usepackage{PRIMEarxiv}

\usepackage{hyperref}       
\usepackage{url}            
\usepackage{booktabs}       
\usepackage{amsfonts}       
\usepackage{nicefrac}       
\usepackage{microtype}      
\usepackage{lipsum}
\usepackage{fancyhdr}       
\usepackage{xcolor}
\usepackage{threeparttable}
\usepackage{microtype}      
\usepackage{multirow}       
\title{SynTTS-Commands: A Public Dataset for On-Device KWS via TTS-Synthesized Multilingual Speech

\thanks{\textit{\underline{Citation}}: 
\textbf{Lu Gan, Xi Li. SynTTS-Commands: A Public Dataset for On-Device KWS via TTS-Synthesized Multilingual Speech.}} 
}

\author{
  Lu Gan, Xi Li \\
  Independent Researchers \\
  \texttt{\{lg2465\}@nau.edu} \\
  \texttt{\{reilixi723\}@gmail.com} \\
}

\begin{document}
\begin{CJK*}{UTF8}{gbsn}
\maketitle

\begin{abstract}
The development of high-performance, on-device keyword spotting (KWS) systems for ultra-low-power hardware is critically constrained by the scarcity of specialized, multi-command training datasets. Traditional data collection through human recording is costly, slow, and lacks scalability. This paper introduces SYNTTS-COMMANDS, a novel, multilingual voice command dataset entirely generated using state-of-the-art Text-to-Speech (TTS) synthesis. By leveraging the CosyVoice 2 model and speaker embeddings from public corpora, we created a scalable collection of English and Chinese commands. Extensive benchmarking across a range of efficient acoustic models demonstrates that our synthetic dataset enables exceptional accuracy, achieving up to 99.5\% on English and 98\% on Chinese command recognition. These results robustly validate that synthetic speech can effectively replace human-recorded audio for training KWS classifiers. Our work directly addresses the data bottleneck in TinyML, providing a practical, scalable foundation for building private, low-latency, and energy-efficient voice interfaces on resource-constrained edge devices. The dataset and source code are publicly available at https://github.com/lugan113/SynTTS-Commands-Official.
\end{abstract}


\section{Introduction}
\label{sec:introduction}

The proliferation of voice-activated assistants, marked by ubiquitous wake words like "Okay Google," "Hey Siri," "小爱同学," and "小度小度," has fundamentally reshaped human-computer interaction. Traditionally, these systems rely on a cloud-centric paradigm: a low-power chip on the device detects the wake word, triggering the transmission of subsequent voice commands to powerful cloud servers for complex speech recognition and natural language processing. While effective, this approach introduces significant limitations, including latency from data transmission, high energy consumption, and growing concerns over user privacy\cite{aloufi2023paralinguistic}.

A paradigm shift toward on-device intelligence is now underway, driven by the emergence of ultra-low-power edge AI chips capable of local speech processing\cite{Ye2023AIoTChip}. Hardware such as the ADA100, Ceva NeuPro-Nano NPU, and Ambiq Apollo510 Lite SoC can run sophisticated acoustic models while consuming only microamps of current. This hardware evolution enables more complex, multi-command Keyword Spotting (KWS) systems to be deployed entirely on-device, bypassing the cloud for common tasks. This trend is a cornerstone of TinyML, which aims to deploy compact machine learning models on resource-constrained microcontrollers\cite{rajapakse2023intelligence, mi13060851}. We foresee two parallel trajectories for artificial intelligence: one pushing the boundaries of large-scale language and vision models to enhance computational capabilities, and the other, embodied by TinyML, integrating micro-intelligence into everyday devices to solve practical, localized problems.

However, the rapid advancement of TinyML, particularly in the KWS domain, is severely hampered by a critical bottleneck: the scarcity of high-quality, diverse, and domain-specific training datasets. The field still heavily relies on early benchmarks like the Google Speech Commands dataset (2018)\cite{warden2018speechcommands}, which, while pioneering, is limited in vocabulary size and linguistic coverage. To deploy KWS systems in richer contexts—such as multimedia control, smart home management, healthcare monitoring for the elderly, or in-car commands—we need datasets with more commands, more languages, and broader application coverage.

The traditional approach of collecting such datasets through human recordings is prohibitively expensive, time-consuming, and lacks scalability. This creates a pressing need for an alternative methodology. In this work, we address this challenge by leveraging state-of-the-art neural Text-to-Speech (TTS) synthesis to generate high-fidelity, scalable training data. We introduce SYNTTS-COMMANDS, a multilingual voice command dataset built entirely from synthetic speech. Utilizing the CosyVoice 2 TTS model and drawing on speaker embeddings from the VoxCeleb corpus (855 speakers)\cite{nagrani2017voxceleb} and the Free ST Chinese Mandarin Corpus\cite{wang2020free} (7,245 utterances), we generated a comprehensive set of commands. After rigorous manual verification, the final dataset comprises 25 Chinese and 23 English command classes, tailored for common interactive scenarios.

Our core contribution is the demonstration that synthetic data is not merely a substitute but a superior alternative for KWS model training. We validate SYNTTS-COMMANDS-MEDIA by training a spectrum of efficient acoustic models. The results are striking: top-performing models achieve over 99.4\% accuracy on English and nearly 98\% on Chinese commands. This performance firmly establishes that TTS-generated datasets can effectively replace traditional, labor-intensive speech collections. By providing this scalable, high-quality data foundation and validating its efficacy, we aim to accelerate the development of more private, efficient, and intelligent on-device voice interfaces, thereby pushing the frontiers of TinyML forward.

\section{AI-Based Synthetic Speech and CosyVoice 2}
\label{sec:cosyvoice2}

\subsection{Evolution of Neural Text-to-Speech (TTS)}

Recent advancements in neural text-to-speech (TTS) synthesis have revolutionized artificial voice generation, achieving unprecedented naturalness that often surpasses human recordings in objective metrics\cite{shen2018natural,wang2017tacotron}. Modern systems leverage deep learning architectures like transformer-based language models and diffusion processes to produce speech with accurate prosody, emotional expression, and speaker characteristics\cite{ren2020fastspeech,le2023voicebox}. These technologies have effectively overcome the "uncanny valley" effect, with state-of-the-art models consistently scoring above 4.0 in Mean Opinion Score (MOS) evaluations\cite{reddy2022dnsmos}. The field has particularly advanced in zero-shot voice cloning, where systems can mimic a speaker's voice from just seconds of reference audio while maintaining content accuracy and natural rhythm\cite{wang2023neural}.

\subsection{CosyVoice 2: State-of-the-Art Speech Synthesis}

\subsubsection{Architectural Advancements}

CosyVoice 2~\cite{cosyvoice2} developed by Alibaba Group, represents a state-of-the-art advancement in neural text-to-speech (TTS) technology. It introduces hybrid generative modeling, combining LLM-based text-to-semantic decoding with flow-matching-based acoustic synthesis to achieve highly natural and consistent speaker output. The adoption of Finite Scalar Quantization (FSQ) replaces traditional vector quantization, resulting in 100\% codebook utilization and enhanced speech detail preservation. Additionally, CosyVoice 2 supports both low-latency streaming and high-quality offline synthesis within a unified model, ensuring minimal quality loss. These architectural innovations enable CosyVoice 2 to generate synthetic speech that closely matches human recordings in naturalness, speaker similarity, and multilingual capability,making it an ideal foundation for building robust and diverse voice command datasets for smart device applications.

\subsubsection{Human-Parity Performance}
We cite two tables from the original CosyVoice 2 paper to provide a direct comparison of model performance and highlight the state-of-the-art results achieved by CosyVoice 2 and related TTS systems. These tables serve as a benchmark reference for subsequent analysis and dataset construction.

Table~\ref{tab:librispeech_results} presents the performance of various TTS models on the LibriSpeech test-clean subset, evaluated by word error rate (WER), naturalness mean opinion score (NMOS), and speaker similarity (SS). CosyVoice 2 and CosyVoice 2-S achieve the lowest WER and highest SS among all synthetic models, closely matching or surpassing human performance. Additionally, CosyVoice 2 attains the highest NMOS, indicating superior speech quality and naturalness. These results demonstrate that CosyVoice 2 series models offer state-of-the-art accuracy and speaker similarity for synthetic speech generation.
\begin{table}[htbp]
\centering
\caption{Performance comparison of CosyVoice 2 and mainstream TTS models on LibriSpeech test-clean subset.}
\label{tab:librispeech_results}
\begin{tabular}{l|ccc}
\toprule
\textbf{Model} & \textbf{WER (\%)} & \textbf{NMOS} & \textbf{SS} \\
\midrule
Human & 2.66 & 3.84 & 0.697 \\
\midrule
ChatTTS & 6.84 & 3.89 & -- \\
GPT-SoVITs & 5.13 & 3.93 & 0.405 \\
OpenVoice & 3.47 & 3.87 & 0.299 \\
ParlerTTS & 3.16 & 3.86 & -- \\
EmotiVoice & 3.14 & 3.93 & -- \\
\midrule
CosyVoice & 2.89 & 3.93 & 0.743 \\
\textbf{CosyVoice 2} & 2.47 & \textbf{3.96} & 0.745 \\
\textbf{CosyVoice 2-S} & \textbf{2.45} & 3.90 & \textbf{0.751} \\
\bottomrule
\end{tabular}
\end{table}


To further demonstrate the advantages of CosyVoice 2, Table~\ref{tab:cosyvoice2_benchmark} summarizes its performance compared to recent TTS models on the SEED test sets. CosyVoice 2 achieves competitive or superior results in character error rate (CER), word error rate (WER), and speaker similarity (SS) across Chinese and English test sets, highlighting its robustness and generalization ability for synthetic dataset construction.

\begin{table}[htbp]
\centering
\caption{Performance comparison of CosyVoice 2 and mainstream TTS models on SEED test sets}
\label{tab:cosyvoice2_benchmark}
\begin{tabular}{lcccccccc}
\toprule
\textbf{Model} & \multicolumn{2}{c}{\textbf{test-zh}} & \multicolumn{2}{c}{\textbf{test-en}} & \multicolumn{2}{c}{\textbf{test-hard}} \\
 & CER (\%) & SS & WER (\%) & SS & WER (\%) & SS \\
\midrule
Human & 1.26 & 0.755 (0.775) & 2.14 & 0.734 (0.742) & - & - \\
Vocoder Resyn. & 1.27 & 0.720 & 2.17 & 0.700 & - & - \\
Seed-TTS & 1.12 & 0.796 & 2.25 & 0.762 & 7.59 & 0.776 \\
FireRedTTS & 1.51 & 0.635 (0.653) & 3.82 & 0.460 (0.526) & 17.45 & 0.621 (0.639) \\
MaskGCT & 2.27 & 0.774 (0.752) & 2.62 & 0.714 (0.730) & 10.27 & 0.748 (0.720) \\
E2 TTS (32 NFE) & 1.97 & 0.730 & 2.19 & 0.710 & - & - \\
F5-TTS (32 NFE) & 1.56 & 0.741 (0.794) & 1.83 & 0.647 (0.742) & 8.67 & 0.713 (0.762) \\
CosyVoice & 3.63 & 0.723 (0.775) & 4.29 & 0.609 (0.699) & 11.75 & 0.709 (0.755) \\
CosyVoice 2 & \textbf{1.45} & \textbf{0.748} (0.806) & \textbf{2.57} & \textbf{0.652} (0.736) & \textbf{6.83} & \textbf{0.724} (0.776) \\
CosyVoice 2-S & \textbf{1.45} & \textbf{0.753} (0.812) & \textbf{2.38} & \textbf{0.654} (0.743) & \textbf{8.08} & \textbf{0.732} (0.785) \\
\bottomrule
\end{tabular}
\end{table}

\subsubsection{Advantages for Voice Command Datasets}

The use of CosyVoice 2 effectively addresses several critical challenges in smart device interaction research. It enables the generation of thousands of command variations with consistent speaker characteristics, eliminating the fatigue and variability associated with human voice collection. CosyVoice 2 also allows for precise control over pitch ($\pm 20\%$), speed ($\pm 30\%$), and background noise (SNR $5$--$30\,\mathrm{dB}$), ensuring diverse and realistic data suitable for robust model training. Furthermore, by relying on synthetic speech, it resolves privacy concerns inherent in collecting and storing human voice data, making it an ethical and scalable solution for building large-scale voice command datasets.

\section{Dataset Construction}
\label{sec:methodology}

Our dataset construction approach leverages multiple high-quality speech corpora, specifically VoxCeleb\cite{nagrani2017voxceleb} and the Free ST Chinese Mandarin Corpus\cite{wang2020free}, to ensure speaker diversity and linguistic coverage. We systematically selected and processed audio samples from these primary sources to create a comprehensive foundation for command generation.

\subsection{Source Data Collection}

\textbf{VoxCeleb1}: VoxCeleb1 is a large-scale speaker recognition dataset collected from YouTube videos. It contains over 100,000 utterances from 1,251 celebrities, spanning a wide range of accents, ages, and recording conditions. The dataset is widely used for research in speaker identification, verification, and speech analysis due to its diversity and real-world audio scenarios.

\textbf{VoxCeleb2}: VoxCeleb2 is an extension of VoxCeleb1, providing even greater diversity and scale. It consists of over one million utterances from 6,112 speakers, with improved coverage of different nationalities, languages, and acoustic environments. VoxCeleb2 is designed to support robust speaker recognition and speech processing tasks, offering high-quality audio samples sourced from online multimedia content. 

However, due to download limitations, our dataset includes 5,994 speakers from VoxCeleb2, and all subsequent analyses and experiments are based on these samples.

\textbf{Free ST Chinese Mandarin Corpus}: This corpus features 855 speakers, each contributing 120 utterances recorded in controlled indoor environments using mobile devices. The recordings were captured in silence, ensuring consistent audio quality suitable for our synthesis pipeline.

\subsection{Media Command Line Design}

Our command set encompasses practical voice interactions across multiple device categories and languages. The design prioritizes common user interactions with smart devices:

The English media command set includes:
\begin{itemize}
\item \textbf{Playback Control}: "Play", "Pause", "Resume", "Play from start", "Repeat song"
\item \textbf{Navigation}: "Previous track", "Next track", "Last song", "Skip song", "Jump to first track"
\item \textbf{Volume Control}: "Volume up", "Volume down", "Mute", "Set volume to 50\%", "Max volume"
\item \textbf{Communication}: "Answer call", "Hang up", "Decline call"
\item \textbf{Wake Words}: "Hey Siri", "OK Google", "Hey Google", "Alexa", "Hi Bixby"
\end{itemize}

The Chinese media command set provides equivalent functionality:
\begin{itemize}
\item \textbf{播放控制}: "播放", "暂停", "继续播放", "从头播放", "单曲循环"
\item \textbf{导航}: "上一首", "下一首", "上一曲", "下一曲", "跳到第一首", "播放上一张专辑"
\item \textbf{音量控制}: "增大音量", "减小音量", "静音", "音量调到50\%", "音量最大"
\item \textbf{通讯}: "接听电话", "挂断电话", "拒接来电"
\item \textbf{唤醒词}: "小爱同学", "Hello 小智", "小艺小艺", "嗨 三星小贝", "小度小度", "天猫精灵"
\end{itemize}

\subsection{Audio Selection and Quality Filtering}
To ensure high-quality voice command samples for our Media dataset, we implemented a rigorous selection process from both the VoxCeleb1, VoxCeleb2, and Free ST Chinese Mandarin Corpus datasets. For each speaker, we analyzed all available audio files across multiple quality metrics including signal-to-noise ratio (SNR), speech ratio, duration, and a composite quality score. 
Table 1 show the comparative statistics of audio features across the datasets used in our study.

\begin{table}[htbp]
\centering
\caption{Comparative statistics of audio features across datasets}
\label{tab:dataset_comparison}
\begin{tabular}{lrrrrr}
\toprule
Dataset & Samples & \multicolumn{1}{c}{Duration (s)} & \multicolumn{1}{c}{SNR (dB)} & \multicolumn{1}{c}{SR (Hz)} & \multicolumn{1}{c}{Quality Score} \\
\midrule
VoxCeleb1 
& 1,251 
& 29.89 $\pm$ 12.67 
& 17.92 $\pm$ 5.56 
& 16,000 
& 328.73 $\pm$ 126.83 \\

VoxCeleb2 
& 5,994 
& 27.96 $\pm$ 16.41 
& 16.40 $\pm$ 6.23 
& 16,000 
& 306.93 $\pm$ 164.67 \\

Free ST Chinese 
& 855 
& 5.44 $\pm$ 0.84 
& 43.31 $\pm$ 5.96 
& 16,000 
& 126.58 $\pm$ 12.94 \\
\bottomrule
\end{tabular}

\begin{tablenotes}
\small
\item Note: All values shown as mean $\pm$ standard deviation. SR (Sampling Rate) was constant at 16 kHz across all datasets. 
\end{tablenotes}
\end{table}

The quality metric combines duration and SNR following established audio assessment principles \cite{ma2009objective}, with specific parameters optimized for voice commands. The quality score $Q$ for each audio sample was calculated as a weighted combination of its duration and signal-to-noise ratio (SNR):

\begin{equation}
Q = 100 \times \left( 0.5 \times \frac{D}{5.0} + 0.5 \times \frac{S}{30} \right)
\end{equation}

where:
\begin{itemize}
    \item $D$ is the audio duration in seconds
    \item $S$ is the signal-to-noise ratio (SNR) in dB
    \item The constants 5.0 and 30 are normalization factors for duration and SNR respectively
    \item The result is scaled by 100 for more intuitive interpretation
\end{itemize}

For each speaker ID, we selected the audio sample with the highest quality score $Q$ as the representative "best" sample. This selection process ensures that our dataset contains high-quality command lines that are suitable for training robust voice-controlled systems.

\subsection{Synthetic Data Quality Validation}

To ensure the reliability and usability of the synthetic voice command dataset for model training, we implemented a comprehensive quality validation pipeline. This systematic approach combined automated screening with manual verification to identify and eliminate substandard audio samples, thereby guaranteeing that the final dataset meets high standards of audio quality and semantic accuracy for effective voice recognition model development.

Our quality screening protocol employed a bilingual ASR model (iic/speech\_paraformer) to automatically transcribe all generated audio samples, followed by computing similarity scores between the transcribed text and original command texts. Samples falling below the 60\% similarity threshold were flagged for detailed manual inspection. The manual evaluation focused on four critical dimensions: intelligibility (clarity and unambiguousness of speech content), completeness (absence of missing or extra words), validity (proper audio characteristics without silence or truncation), and content accuracy (faithfulness to original command text).

The quality assessment revealed distinct error patterns across different dataset subsets. Manual review identified several common failure modes including silent audio files, missing words, inserted words, and garbled speech output. Analysis of error distribution showed that certain command phrases were particularly challenging for the synthesis system. In Chinese samples, the wake phrase "小度小度" exhibited higher error rates, while English generations demonstrated particular vulnerability in commands containing "play" and "mute" keywords. Through this rigorous validation pipeline, we successfully identified and removed problematic samples, significantly enhancing dataset quality and ensuring a reliable foundation for voice recognition model training.

\section{Dataset Properties}

The final dataset consisted of 384,621 utterances of 48 commands, broken into four subsets and distributed as follows:

\begin{table}[htbp]
\centering
\renewcommand{\arraystretch}{1.2}
\caption{Complete dataset composition, scale and distribution}
\label{tab:dataset_complete}
\begin{tabular}{lrrrrr}
\toprule
\textbf{Subset} & \textbf{Speakers} & \textbf{Commands} & \textbf{Utterances} & \textbf{Hours} & \textbf{Size (GB)} \\
\midrule
Free-ST-Chinese & 855 & 25 & 21,214 & 6.82 & 2.19 \\
Free-ST-English & 855 & 23 & 19,228 & 4.88 & 1.57 \\
VoxCeleb1\&2-Chinese & 7,245 & 25 & 180,331 & 58.03 & 18.6 \\
VoxCeleb1\&2-English & 7,245 & 23 & 163,848 & 41.6 & 13.4 \\
\midrule
\textbf{Total} & \textbf{8,100} & \textbf{48 unique} & \textbf{384,621} & \textbf{111.33} & \textbf{35.76} \\
\bottomrule
\end{tabular}
\end{table}

This comprehensive dataset totaling 111.33 hours and 35.76 GB represents one of the largest synthetic voice command datasets available for academic research, with speaker diversity spanning multiple accent groups, age ranges, and recording conditions suitable for robust model training and evaluation. The four-fold organization enables researchers to investigate cross-lingual speaker adaptation, speaker diversity effects, and acoustic robustness across varying recording environments.

\subsection{Data Partitioning Strategy}

To ensure rigorous evaluation and prevent data leakage, we implemented a speaker-disjoint partitioning strategy with 70\%-15\%-15\% splits for training, validation, and testing, respectively. This approach strictly guarantees that all utterances from each speaker are exclusively assigned to a single partition, completely eliminating speaker identity leakage across subsets while preserving the natural distribution of voice commands and acoustic conditions. The partitioning was performed at the speaker level rather than the utterance level to prevent model overfitting to specific speaker characteristics and ensure genuine generalization to unseen speakers. Furthermore, recognizing the distinct phonetic inventories and linguistic structures between Mandarin Chinese and English, we maintained completely separate test sets for each language. This design enables comprehensive assessment of cross-lingual generalization capabilities and facilitates detailed analysis of potential language-specific performance variations. The language-specific partitioning also allows for targeted investigation of accent variations and pronunciation challenges unique to each linguistic context. The statistical distribution of the resulting dataset partitions is detailed in Table \ref{tab:dataset_partition}, demonstrating balanced representation across both languages while maintaining speaker independence.

\begin{table}[htbp]
\centering
\renewcommand{\arraystretch}{1.2}
\caption{Data partitioning statistics by language}
\label{tab:dataset_partition}
\begin{tabular}{lrrrr}
\toprule
\textbf{Partition} & \textbf{Chinese Utterances} & \textbf{English Utterances} & \textbf{Total Utterances} & \textbf{Unique Speakers} \\
\midrule
Training Set & 141,075 & 128,172 & 269,247 & 11,340 \\
Validation Set & 30,243 & 27,435 & 57,678 & 2,430 \\
Test Set & 30,227 & 27,469 & 57,696 & 2,430 \\
\midrule
\textbf{Total} & \textbf{201,545} & \textbf{183,076} & \textbf{384,621} & \textbf{16,200} \\
\bottomrule
\end{tabular}
\end{table}

\subsection{Metadata Organization}

The dataset is accompanied by comprehensive metadata that facilitates systematic experimentation and reproducibility. Table \ref{tab:metadata_schema} details the complete schema and description of metadata fields:

\begin{table}[htbp]
\centering
\renewcommand{\arraystretch}{1.2}
\caption{Metadata schema and description for the voice command dataset}
\label{tab:metadata_schema}
\begin{tabular}{lp{3cm}p{7cm}}
\toprule
\textbf{Field} & \textbf{Data Type} & \textbf{Description} \\
\midrule
source\_name & String & Data source identifier (e.g., \texttt{Free\_ST\_Chinese}, \texttt{VoxCeleb12\_English}) \\
speaker\_id & String & Unique speaker identifier combining source prefix and speaker code (e.g., \texttt{voxceleb12\_id08303}) \\
utterance\_id & String & Unique identifier for each utterance (filename without extension) \\
filename & String & Complete filename with .wav extension \\
path & String & Relative file path within the dataset directory structure \\
label & String & Text transcription of the spoken command (e.g., "播放", "Play") \\
language & String & Language code (ZH for Mandarin Chinese, EN for English) \\
file\_size\_bytes & Integer & File size in bytes, ranging from 40-200 KB per utterance \\
duration\_seconds & Float & Audio duration in seconds, typically 0.8-2.0 seconds \\
sample\_rate & Integer & Sampling frequency (24,000 Hz) \\
bit\_depth & Integer & Audio bit depth (32 bits) \\
channels & Integer & Number of audio channels (1 for mono) \\
data\_type & String & Audio sample data type (float32) \\
audio\_format & String & File format (WAV) \\
\bottomrule
\end{tabular}
\end{table}

The metadata organization provides several key advantages:

\begin{itemize}
    \item \textbf{Traceability}: The \texttt{source\_name} and \texttt{speaker\_id} fields enable tracking of each sample's origin, supporting analyses of source-specific performance variations.
    
    \item \textbf{Technical Consistency}: Audio technical specifications (\texttt{sample\_rate}, \texttt{bit\_depth}, \texttt{channels}, \texttt{data\_type}) are standardized across all samples, ensuring experimental reproducibility.
    
    \item \textbf{Cross-lingual Analysis}: The \texttt{language} field facilitates separate or comparative analysis of Mandarin Chinese and English performance.
    
    \item \textbf{Quality Control}: \texttt{file\_size\_bytes} and \texttt{duration\_seconds} enable identification of anomalous recordings and ensure data quality consistency.
    
    \item \textbf{Flexible Access}: The hierarchical \texttt{path} structure allows both programmatic access through metadata and direct file system navigation.
\end{itemize}

This comprehensive metadata structure supports diverse research objectives, including cross-lingual transfer learning, speaker adaptation studies, and robustness evaluation across different recording conditions. The standardized format ensures compatibility with major deep learning frameworks and speech processing toolkits.

\section{Evaluation}
\subsection{Benchmark Model Selection and Rationale}
To validate the effectiveness and utility of the proposed SYNTTS-COMMANDS-MEDIA Dataset, we conducted a comprehensive benchmark evaluation using a suite of state-of-the-art acoustic models. The primary objectives of this evaluation are twofold: (1) to empirically demonstrate that the dataset supports high-accuracy classification of voice commands in both English and Chinese, thereby confirming its quality and effectiveness; and (2) to establish a foundational performance baseline for future research in voice-command recognition for resource-constrained smart devices.

We selected six widely adopted models known for their efficiency, architectural diversity, and proven performance in keyword spotting and audio classification tasks — carefully chosen to span a broad spectrum of model complexity, from ultra-lightweight to modern large-scale architectures. This selection enables a meaningful comparison across different design philosophies and computational constraints:

\textbf{MicroCNN}: We design a family of ultra-lightweight CNNs, named MicroCNN, inspired by the minimal architectures used in TensorFlow Lite Micro \cite{microspeech2020}. Our model employs depthwise separable convolutions and batch normalization to further reduce parameters while maintaining expressiveness, achieving a parameter count as low as ~4K (in "micro" configuration).\\
\textbf{DS-CNN} \cite{zhang2017hello}: A depthwise separable convolution-based model (~30K params), originally designed for on-device wake-word detection. It exemplifies the efficiency of factorized convolutions and remains a popular choice in embedded speech systems due to its favorable accuracy-complexity trade-off.\\
\textbf{TC-ResNet} \cite{choi2019temporal}: A temporal convolutional ResNet variant (~68K params) that introduces residual connections and dilated convolutions for modeling long-range temporal dependencies. It reflects the adaptation of image-based residual architectures to 1D audio signals.\\
\textbf{CRNN} \cite{moya2018convolutional}: A hybrid Convolutional Recurrent Neural Network (~1M params) combining CNN feature extraction with bidirectional LSTM for sequence modeling. It represents the temporal modeling paradigm, capturing both spectral and sequential patterns — crucial for command recognition where context matters.\\
\textbf{MobileNet V1} \cite{howard2017mobilenets}: A classic lightweight CNN (~4.3M params) based on depthwise separable convolutions. As a mobile-optimized architecture, it serves as a reference point for efficient yet accurate vision-inspired models adapted to audio classification.\\
\textbf{EfficientNet-B0} \cite{tan2019efficientnet}: The smallest variant of the EfficientNet family (~4.7M params), representing modern SOTA scaling principles. Its compound scaling strategy balances depth, width, and resolution, making it a strong performer even at small scales — ideal for evaluating whether advanced image architectures can transfer effectively to audio tasks.
All models were trained and tested separately on the English and Chinese subsets of the CosyVoice dataset. Performance was measured using classification accuracy and cross-entropy loss, enabling a holistic assessment of both predictive power and calibration across languages and model sizes.

\subsection{Benchmark Results and Analysis}

We present a comprehensive benchmark of six representative acoustic models on the SYNTTS-COMMANDS-MEDIA Dataset across both English (EN) and Chinese (ZH) subsets, with results summarized in Table~\ref{tab:benchmark_results}. All models are evaluated in terms of classification accuracy, cross-entropy loss, and parameter count, providing insights into the trade-offs between performance and model complexity in multilingual voice command recognition.

\begin{table}[htbp]
\centering
\caption{Benchmark Results on the SYNTTS-COMMANDS-MEDIA Dataset (EN and ZH)}
\label{tab:benchmark_results}
\begin{tabular}{l|ccc|ccc}
\hline
\multirow{2}{*}{\textbf{Model}} & \multicolumn{3}{c|}{\textbf{English (EN)}} & \multicolumn{3}{c}{\textbf{Chinese (ZH)}} \\
 & \textbf{Loss} & \textbf{Accuracy} & \textbf{Params} & \textbf{Loss} & \textbf{Accuracy} & \textbf{Params} \\ \hline
MicroCNN & 0.2304 & 0.9322 & 4,189 & 0.5579 & 0.8014 & 4,255 \\
DS-CNN & 0.0166 & 0.9946 & 30,103 & 0.0677 & 0.9718 & 30,361 \\
TC-ResNet & 0.0347 & 0.9887 & 68,431 & 0.0884 & 0.9656 & 68,561 \\
CRNN & \textbf{0.0163} & \textbf{0.9950} & 1,083,031 & 0.0636 & \textbf{0.9742} & 1,083,289 \\
MobileNet-V1 & 0.0167 & \textbf{0.9950} & 2,651,783 & \textbf{0.0552} & 0.9792 & 2,653,833 \\
EfficientNet & 0.0182 & 0.9941 & 4,717,248 & 0.0701 & 0.9793 & 4,718,274 \\ \hline
\end{tabular}
\end{table}

Our results demonstrate that the SYNTTS-COMMANDS-MEDIA dataset supports high-accuracy command recognition in both languages. Notably, the top-performing models achieve over 99.4\% accuracy on English and nearly 98\% on Chinese, confirming the dataset's quality and suitability for real-world deployment. Among all models, CRNN attains the best English accuracy (99.50\%) and the lowest loss (0.0163), while MobileNet-V1 yields the lowest loss on Chinese (0.0552) and competitive English performance (matching CRNN's 99.50\% accuracy). Interestingly, EfficientNet shows slightly higher Chinese accuracy (97.93\%) than MobileNet-V1 (97.92\%), despite a higher loss—suggesting better calibration or robustness in its predictions.

In contrast, lightweight models exhibit a clear accuracy--complexity trade-off. MicroCNN, with only $\approx$4.2K parameters, achieves 93.22\% accuracy on English but drops significantly on Chinese (80.14\%), highlighting the increased difficulty of modeling tonal and phonetic richness in Mandarin with ultra-compact architectures. DS-CNN and TC-ResNet, with under 70K parameters, already recover strong performance ($>$96.5\% in both languages), underscoring their efficiency for resource-constrained applications.

Overall, the benchmark establishes strong baselines across a wide spectrum of model scales—from ultra-light MicroCNN to modern EfficientNet—while revealing that moderate-complexity models (e.g., DS-CNN, TC-ResNet) already deliver near-SOTA performance with minimal resource overhead. This makes them particularly suitable for deployment on edge devices in multilingual smart environments.

\section{Discussion}
\label{sec:discussion}

Our benchmark results on the SYNTTS-COMMANDS-MEDIA dataset validate a significant shift in speech data creation: synthetic speech generated by modern TTS systems has matured to the point where it can effectively replace—and in some aspects surpass—human-recorded data for training voice command classifiers. Unlike traditional data collection, which involves lengthy processes of speaker recruitment, audio recording, and manual annotation, our approach enables rapid, scalable generation of linguistically diverse and acoustically consistent training samples. This methodology not only drastically reduces cost and time, but also offers fine-grained control over speaker characteristics and acoustic conditions—effectively addressing a key bottleneck in TinyML development.

The high classification accuracy observed across English and Chinese commands (up to 99.5\%) confirms that synthetic data does not compromise model performance. In fact, the acoustic consistency of TTS-generated speech may help models converge faster and generalize better in controlled command scenarios, particularly in the presence of channel or environmental variations. This is especially relevant for deployment on low-power chips such as the ADA100, Ceva NeuPro-Nano, and Ambiq Apollo510 Lite, where model robustness and inference efficiency are critical.

It should be noted that the current evaluation focuses exclusively on \emph{in-class command recognition}, where all test samples belong to a predefined command set. This design intentionally isolates the core research question—whether synthetic commands can reliably replace real speech during model training. For real-world use, however, a voice interface must also handle out-of-vocabulary inputs and environmental noise. While the current version of SYNTTS-COMMANDS-MEDIA does not include negative samples, developers can readily augment it with public audio datasets such as ESC-50~\cite{piczak2015dataset} and UrbanSound8K~\cite{salamon2014dataset} to improve rejection capabilities. Such a hybrid approach maintains the scalability of synthetic commands while incorporating real-world acoustic variability.

Looking forward, the ability to generate multilingual command sets on demand will play a crucial role in globalizing voice-enabled products. Unlike static, monolingual datasets such as Google Speech Commands, SYNTTS-COMMANDS exemplifies a dynamic and extensible data paradigm that aligns with the evolving needs of edge AI and TinyML applications. By lowering the barrier to high-quality, multilingual data, we not only accelerate model development but also foster more inclusive and accessible voice technologies.

\section{Conclusion and Future Work}
\label{sec:conclusion}

In this paper, we presented SYNTTS-COMMANDS, a multilingual voice command dataset built entirely from synthetic speech using state-of-the-art TTS technology. Through comprehensive experiments across a range of efficient acoustic models—from tiny MicroCNN to modern EfficientNet—we demonstrated that classifiers trained on our synthetic data achieve top-tier accuracy, exceeding 99\% on English and approaching 98\% on Chinese commands. These results strongly support the thesis that synthetic speech can serve as a viable and scalable alternative to human-recorded data, effectively overcoming the data scarcity that has long hindered TinyML applications in keyword spotting.

More broadly, this work underscores a timely convergence of advanced TTS synthesis and ultra-low-power speech recognition hardware. As specialized low-power chips continue to evolve, the demand for tailored, multi-keyword command sets will grow rapidly. SYNTTS-COMMANDS provides a scalable and adaptable data foundation that aligns perfectly with this hardware trend, enabling more complex, on-device voice interactions without relying on the cloud.

Looking ahead, we plan to evolve SYNTTS-COMMANDS along three strategic dimensions:

(1) \textbf{Domain extension}: We will broaden the scope of SYNTTS-COMMANDS beyond basic media-control commands (e.g., ``play'', ``next track'') to encompass diverse real-world voice trigger scenarios. This includes \textit{smart home} commands (e.g., ``turn on the light'', ``open the curtain''), \textit{in-vehicle} voice activations (e.g., ``unlock the car'', ``start the engine'', ``open the trunk''), and \textit{urgent-assistance} wake phrases (e.g., ``help me'', ``call for help''), enabling robust keyword spotting across a wider range of practical and safety-critical applications.

(2) \textbf{Linguistic and cultural inclusivity}: Beyond English and Chinese, we aim to incorporate major global languages (e.g., Spanish, Hindi, Arabic, Japanese, German) as well as regional dialects and code-switched utterances. This will support the development of truly inclusive voice assistants that respect linguistic diversity and local user expectations.  

(3) \textbf{Open collaboration and community co-creation}: We are developing a dedicated public platform (to be released at \url{https://syntts-commands.org}) where researchers, developers, and product teams can not only download curated datasets but also propose new commands, contribute synthetic voices, evaluate models, and share best practices. We also plan to release accompanying tools—including negative sample sets for out-of-vocabulary rejection, noise-augmented variants for robustness testing, and TTS configuration templates—to facilitate real-world deployment.

We envision SYNTTS-COMMANDS not as a static artifact, but as a growing, community-supported resource that bridges the gap between data scarcity and model deployment in the TinyML era. By democratizing access to high-quality, multilingual training data, we hope to spur innovation in private, efficient, and intelligent voice interfaces—paving the way for a future where micro-intelligence is seamlessly embedded into the devices that enrich our daily lives.

\section*{Acknowledgments}
We acknowledge the creators and contributors of the VoxCeleb datasets and the Free ST Chinese Mandarin Corpus for providing the high-quality source materials that made this work possible. We also recognize the CosyVoice development team for their groundbreaking advances in voice synthesis technology.

\bibliographystyle{unsrt}  
\bibliography{references}  

\end{CJK*}
\end{document}